\documentclass[final,5p,times,twocolumn]{elsarticle}

\usepackage{graphicx}

\usepackage{amssymb}

\usepackage{lineno,hyperref}
\modulolinenumbers[5]

\journal{Nuclear Instruments and Methods A}

\begin{document}

\begin{frontmatter}

\title{Two-phase Cryogenic Avalanche Detector with electroluminescence gap operated in argon doped with nitrogen}

\author[A,B]{A. Bondar}
\author[A,B]{A. Buzulutskov\corref{corresponding author}}
\author[B]{A. Dolgov}
\author[A,B]{V. Nosov}
\author[A,B]{L. Shekhtman}
\author[A,B]{E. Shemyakina}
\author[A,B]{A. Sokolov}
\address[A]{Budker Institute of Nuclear Physics SB RAS, Lavrentiev avenue 11, 630090 Novosibirsk, Russia}
\address[B]{Novosibirsk State University, Pirogov street 2, 630090 Novosibirsk, Russia}

\cortext[corresponding author]{Corresponding author}







\begin{abstract}
A two-phase Cryogenic Avalanche Detector (CRAD) with electroluminescence (EL) gap, operated in argon doped with a minor (49$\pm$7 ppm) admixture of nitrogen, has been studied. The EL gap was optically read out using cryogenic PMTs located on the perimeter of the gap. We present the results of the measurements of the N$_2$ content, detector sensitivity to X-ray-induced signals, EL gap yield and electron lifetime in the liquid. The detector sensitivity, at a drift field in liquid Ar of 0.6 kV/cm, was measured to be 9 and 16 photoelectrons recorded at the PMTs per keV of deposited energy at 23 and 88 keV respectively. Such two-phase detectors, with enhanced sensitivity to the S2 (ionization-induced) signal, are relevant in the field of argon detectors for dark matter search and low energy neutrino detection.
\end{abstract}


\end{frontmatter}


\section{Introduction}

The present study was performed in the course of the development of two-phase Cryogenic Avalanche Detectors (CRADs) \cite{CRADPropEL,RevCRAD}, operated in Ar, for dark matter search \cite{Warp,Darkside} and coherent neutrino-nucleus scattering \cite{CoNu1,CoNu2} experiments and their energy calibration \cite{LArIonYieldBern1,LArIonYieldScene,LArIonYieldCRAD,XRayYield}.

In two-phase detectors operated in Ar, the S2 signal (induced by the primary ionization in the noble-gas liquid), is detected through the effect of proportional electroluminescence (or proportional scintillation) in electroluminescence (EL) gap located directly above the liquid-gas interface \cite{Rev1}. In proportional electroluminescence the energy provided to the electrons by the electric field is almost fully expended in atomic excitations producing $Ar^{\ast}(3p^54s^1)$ and $Ar^{\ast}(3p^54p^1)$ states. These are followed by the photon emission in the Vacuum Ultraviolet (VUV), around 128 nm, due to excimer ($Ar^{\ast}_2$) productions in three-body collisions and their subsequent decays \cite{PropELMech,PropELGAr,PropELSim}, and by the photon emission in the Near Infrared (NIR), at 690-850 nm, due to $Ar^{\ast}(3p^54p^1)\rightarrow Ar^{\ast}(3p^54s^1)+h\nu$ atomic transitions \cite{NirCRAD,NirYieldSim}.

In presence of nitrogen admixture to gaseous argon the mechanism of proportional electroluminescence is modified. Namely, the excimer production (and hence the VUV emission) can be taken over by that of excited N$_2$ molecules in two-body collisions followed by their de-excitations in the Near Ultraviolet (NUV) \cite{PropELArN2}, through the emission of the so-called Second Positive System (SPS), at 260-430 nm \cite{ArN2SPS}.

It should be remarked that VUV recording is rather inefficient in two-phase detectors in Ar as compared to that of NUV, due to re-emission and total reflection losses in the Wavelength Shifter (WLS): the losses in light collection efficiency may exceed an order of magnitude \cite{CRADPropEL}. Accordingly, it looks attractive to dope argon with a small amount of nitrogen, to convert the VUV into the NUV directly in the EL gap. This idea has been recently realized by our group in \cite{CRADPropEL}. A high performance two-phase CRAD has been developed there, with EL gap optically read out using cryogenic PMTs located on the perimeter of the gap, and with combined THGEM/GAPD-matrix  multiplier  \cite{RevCRAD,NirCRAD,CRADMatrix} (THGEM is Thick Gas Electron Multiplier \cite{THGEMRev}; GAPD is Geiger-mode APD or MPPC or else SiPM \cite{CryoGAPD}). Such a combined charge/optical readout of two-phase detectors would result in a higher overall gain at superior spatial resolution.

In that work \cite{CRADPropEL} we systematically studied proportional electroluminescence in Ar in the two-phase mode, with a minor ($\sim$50 ppm) admixture of N$_2$,  that might be typical for large-scale liquid Ar experiments \cite{LArN2Impure,LargeLArRef}. The results obtained there indicate that the effect of N$_2$ doping on proportional electroluminescence in Ar is enhanced at lower temperatures. In addition, the S2 signal response was found to be substantially (by a factor of 3) enhanced due to the fraction of the N$_2$ emission spectrum recorded directly, i.e. escaping re-emission in the WLS film and thus having a considerably higher light collection efficiency.

In this work we present some topics on the performance of the two-phase CRAD with EL gap in N$_2$-doped Ar, not covered in the previous work, namely the results of the measurements of the N$_2$ content, amplitude characteristics for X-ray-induced signals, EL gap yield and electron lifetime in the liquid.

\section{Experimental setup and procedures}

\begin{figure}[hbt]
\centering
\includegraphics[width=\columnwidth,keepaspectratio]{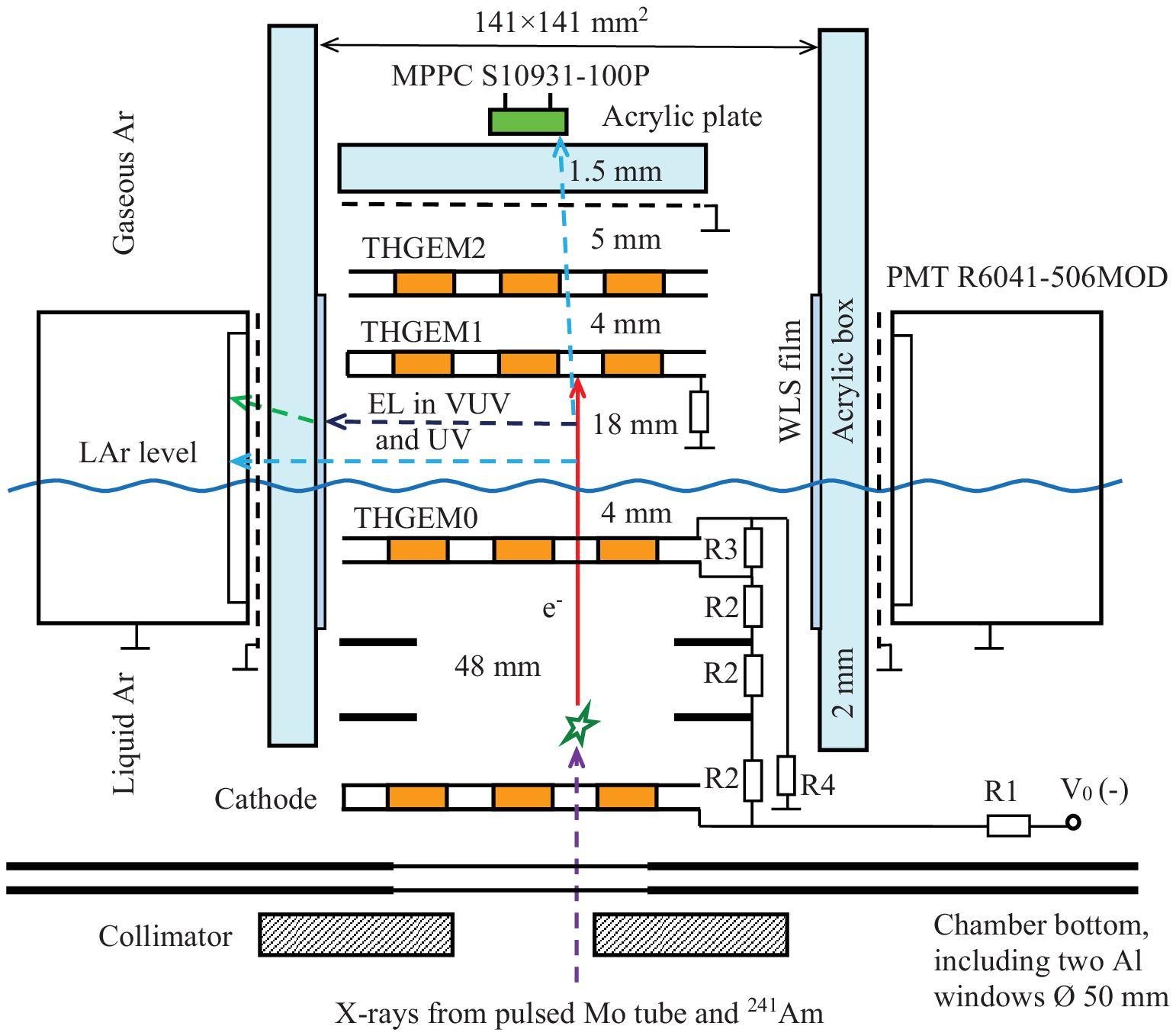}
\caption{Schematic view of the experimental setup (not to scale). The resistors of the voltage divider have the following values: R1, R2, R3 and R4 is 80, 40, 4 and 600 MOhm respectively.}
\label{Setup}
\end{figure}

Fig.~\ref{Setup} shows the experimental setup; it is described elsewhere \cite{CRADPropEL}. Here we present the details relevant to the present study. The setup comprised a 9 l cryogenic chamber filled with 2.5 liters of liquid Ar. Ar was taken from a bottle with a specified purity of 99.998\%. During each cooling procedure Ar was purified from electronegative impurities by Oxisorb filter, providing electron lifetime in the liquid $>$100 $\mu$s. The detector was operated in two-phase mode in the equilibrium state, at a saturated vapor pressure of 1.000$\pm$0.003 atm and at a temperature of 87.3 K.

It should be noted that a reliable measurement of the N$_2$ content in the cryogenic system at ppm level is a challenge; we describe below how this problem was solved. After each cryogenic measurement, the Ar was liquified from the chamber back to a stainless steel bottle ("CRAD bottle") cooled with liquid nitrogen, so that the N$_2$ content remained constant throughout the entire measurement campaign. The latter lasted 5 months during which the setup (operated on a closed loop) was repeatedly evacuated but neither baked nor purified from N$_2$, resulting in a certain N$_2$ content established in the system due to combined effect of the residual gas and internal outgassing in the bottle and cryogenic chamber.  At the end of the campaign, the N$_2$ content in the CRAD bottle  was measured in two ways: first, using a Residual Gas Analyzer (RGA) Pfeiffer-Vacuum QME220, and second, using a chromatography technique.

\begin{figure}[hbt]
\centering
\includegraphics[width=\columnwidth,keepaspectratio]{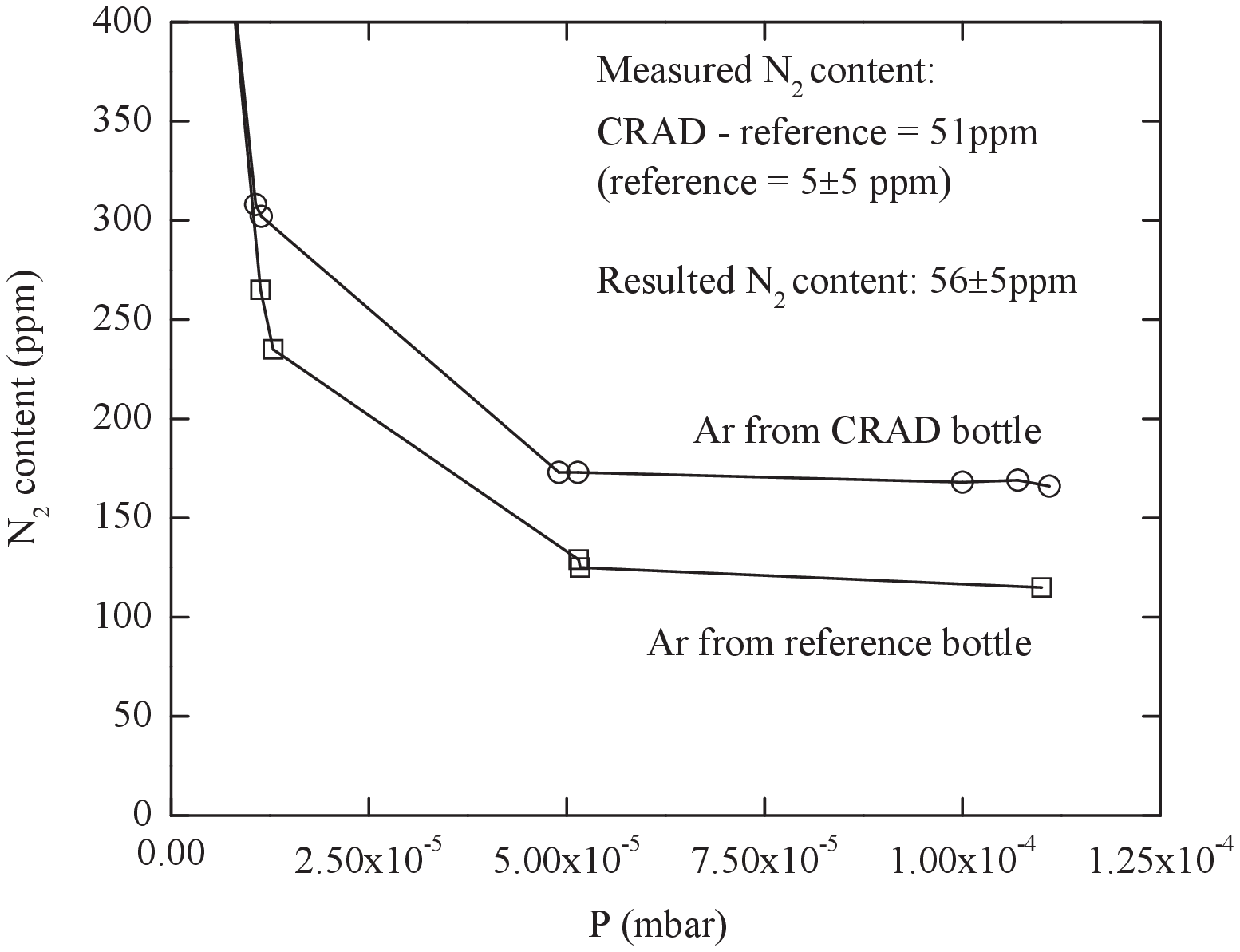}
\caption{N$_2$ content measurement using RGA technique. The N$_2$ content in Ar as measured by the RGA is shown as a function of the residual pressure, for the CRAD and reference bottles.}
\label{N2content}
\end{figure}

In case of RGA technique, the CRAD bottle was connected to a baked high-vacuum ($4\times10^{-9}$ mbar) test chamber equipped with RGA, where the N$_2$ content in Ar was measured in a flow mode at a pressure reaching $10^{-4}$ mbar: see Fig.~\ref{N2content}. It should be remarked that the RGA readings were difficult to get rid of the influence of outgassing from the walls of the test chamber and lead-in tubes. Therefore only the relative N$_2$ content had the meaning, measured with respect to the reference bottle with a known N$_2$ content.  Thus the N$_2$ content value obtained using this technique was 56$\pm$5 ppm: it was composed of the difference between the RGA readings for the CRAD and the reference bottles (51 ppm) and of the N$_2$ content of the reference bottle (5$\pm$5 ppm) specified by the manufacturer.

In case of chromatography technique, the N$_2$ content in the CRAD bottle was measured by a local industrial company "Pure Gases": it amounted to 42 ppm. Thus the final value of N$_2$ content, averaged over the two techniques, amounted to 49$\pm$7 ppm. In the two-phase mode at 87 K, this value corresponds to the N$_2$ content of 49 ppm in the liquid and 132 ppm in the gas phase, according to "Raoult" law \cite{TPArN2CRAD}.

The cryogenic chamber included a cathode electrode, field-shaping electrodes and a THGEM0, immersed in a 55 mm thick liquid Ar layer. These elements were biased through a resistive high-voltage divider placed within the liquid, forming a drift region in liquid Ar, 48 mm long. A 4 mm thick liquid Ar layer above the THGEM0 acted as an electron emission region. A double-THGEM assembly, consisting of a THGEM1 and THGEM2, was placed in the gas phase above the liquid. The EL gap (the EL region), 18 mm thick, was formed by the liquid surface and the THGEM1 plate; the latter was grounded through a resistor acting as an anode of the gap. All electrodes had the same active area, of 10$\times$10 cm$^2$.

The voltage applied to the divider varied from 11 to 22 kV, producing the electric drift field in liquid Ar of 0.34-0.68 kV/cm, electric emission field of 2.6-5.1 kV/cm and electric field in the EL gap of 4.0-8.0 kV/cm. Using such a voltage divider, the THGEM0 was biased in a way to provide the effective transmission of drifting electrons from the drift region to that of electron emission: the electrons drifted successively from a lower to higher electric field region, with a field ratio of about 3 at each step, which in principle should result in electron transmittance through the THGEM0 approaching 100\%. The average drift time of the primary ionization electrons across the drift, emission and EL regions varied from about 25 to 35 $\mu$s, depending on the applied electric fields.

The detector was irradiated from outside by X-rays from a pulsed X-ray tube \cite{XRayYield}, with the energy of 30-40 keV after X-ray filtering \cite{CRADPropEL}, those from $^{109}$Cd source, with the energy of 22-25 and 88 keV, and those from $^{241}$Am source, with the energy of 60 keV.

The EL gap was viewed by four compact cryogenic 2-inch PMTs R6041-506MOD \cite{CryoPMT}, located on the perimeter of the gap. The PMTs were electrically insulated from the gap by a grounded mesh and an acrylic protective box of a rectangular shape. To convert the VUV into the blue light, four wavelength shifter (WLS) films, based on TPB (tetraphenyl-butadiene) in polystyrene matrix, were deposited on the inner box surface facing the EL gap, in front of each PMT. In addition, proportional electroluminescence in the spectral range other than the VUV, i.e. in the NUV (and visible and NIR range, if any) could be recorded using a MPPC (Multi-Pixel Photon Counter) S10931-100P \cite{CryoGAPD}, placed behind the THGEM2.

\begin{figure}[hbt]
\centering
\includegraphics[width=0.8\columnwidth,keepaspectratio]{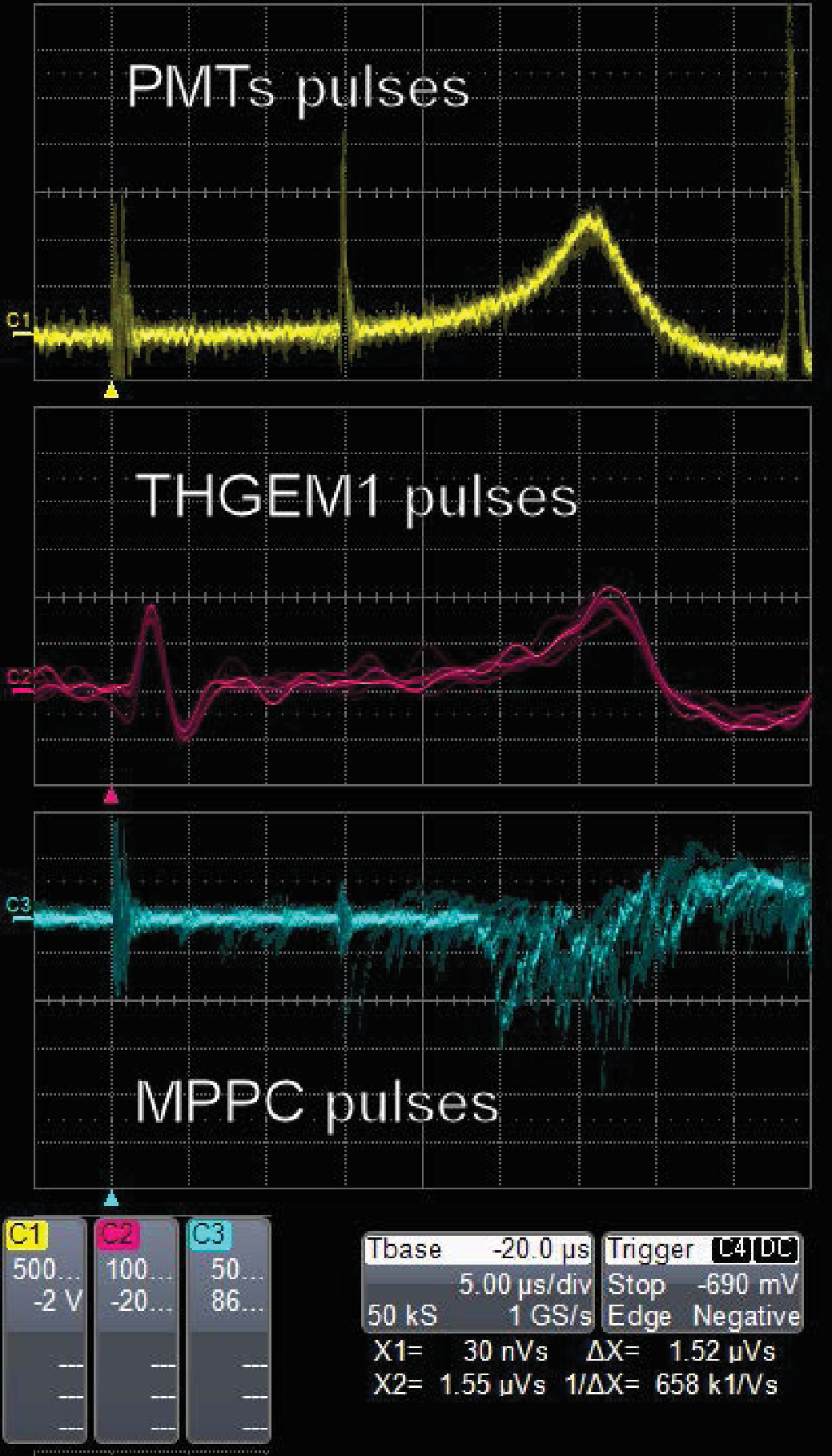}
\caption{Typical signals from the EL gap, namely the total optical signal from the PMTs, the charge signal from the THGEM1 and the optical signal from the MPPC, induced by pulsed X-rays absorbed in liquid Ar, at an electric field of 6.5 kV/cm in the EL gap and 0.55 kV/cm in the drift region. Note that the ionization electron drift time, equal to the distance between the trigger indicated by a triangle and the PMT pulse peak, is about 31 $\mu$s.}
\label{Signals}
\end{figure}

Three types of signals were recorded from the EL gap: the optical signal from the PMTs, the charge signal from the THGEM1 acting as an anode of the gap and the optical signal from the MPPC (see Fig.~\ref{Signals}). The optical signal from the four PMTs, called the total PMT signal, was obtained  as a linear sum of all the PMT signals, amplified with a linear amplifier with a shaping time of 200 ns. The THGEM1 charge signal was recorded using a charge-sensitive preamplifier followed by a shaping amplifier with a time constant of 1 $\mu$s. The MPPC optical signal was recorded using a fast amplifier with a shaping time of 40 ns. Note the rather large drift time of the ionization electrons in Fig.~\ref{Signals}, equal to the distance between the trigger and the PMT pulse peak, which is about 31 $\mu$s.

\section{Amplitude characteristics and EL gap yield}

\begin{figure}[hbt]
\centering
\includegraphics[width=\columnwidth,keepaspectratio]{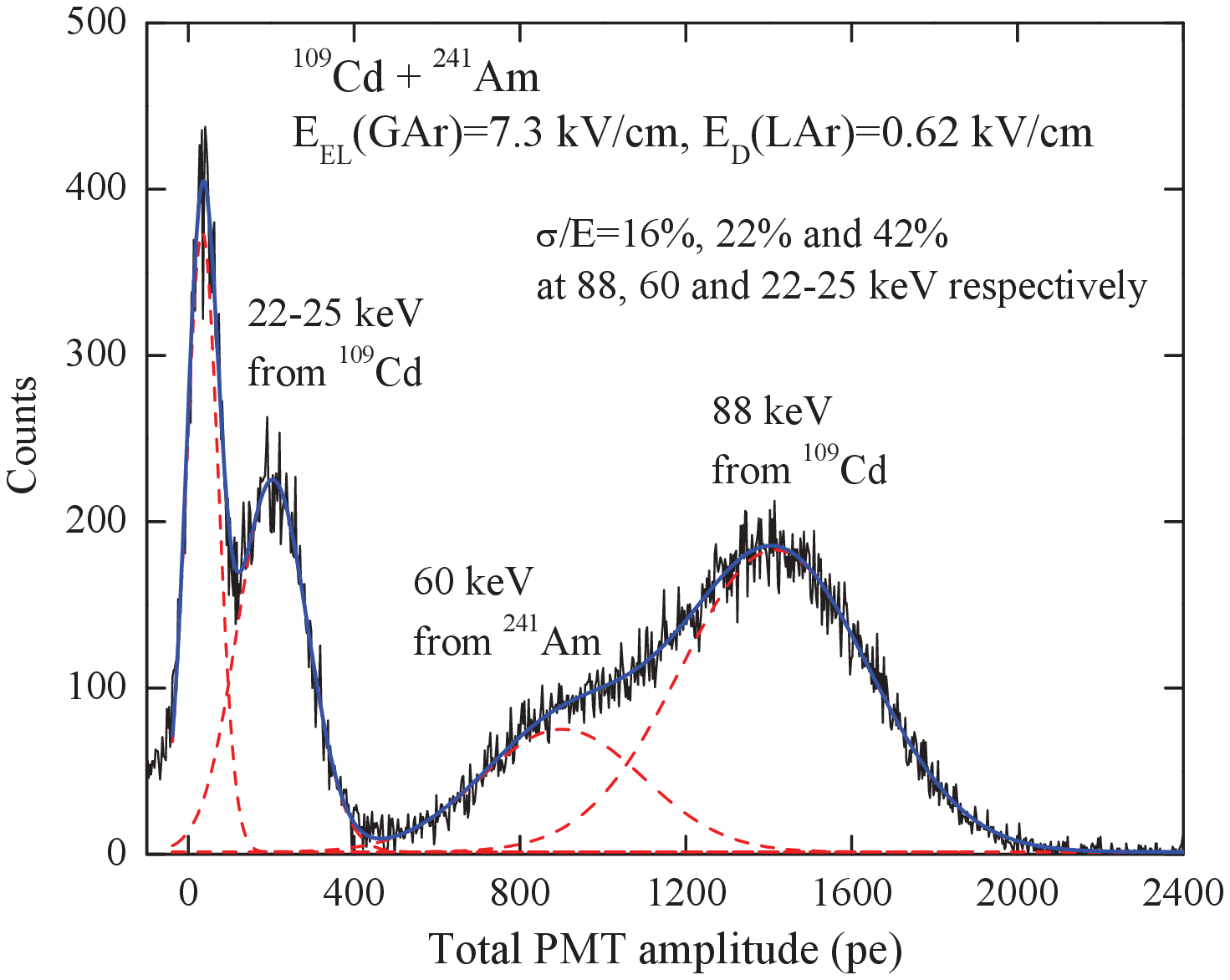}
\caption{Amplitude distribution of the total PMT signal from the EL gap at an electric field of 7.3 kV/cm in the EL gap and 0.62 kV/cm in the drift region, under irradiation with X-rays from $^{109}$Cd and $^{241}$Am radioactive sources. The amplitude is expressed in the number of photoelectrons (pe) recorded at the PMTs.}
\label{CdSpectrum}
\end{figure}

The amplitude characteristics of the detector, when recording the S2 (ionization-induced) signal, are illustrated in Fig.~\ref{CdSpectrum}: it shows the amplitude spectrum of the total PMT signal from the EL gap induced by X-rays from a mixture of the $^{109}$Cd and $^{241}$Am radioactive sources, at an electric field of 7.3 kV/cm in the EL gap and 0.62 kV/cm in the drift region. The peaks due to X-ray lines at 88 keV and 22-25 keV of the $^{109}$Cd source, as well as that at 60 keV of the $^{241}$Am sources, are well distinguished. The energy resolution improved with energy; it amounted to $\sigma/E$=42, 22 and 16\% at 22-25, 59.5 and 88 keV respectively.

The amplitude in Fig.~\ref{CdSpectrum} is expressed in the number of photoelectrons (pe) recorded at the PMTs. This allows one to estimate the detector yield (sensitivity) for electron-equivalent (ee) recoils induced by X-ray absorption: it is defined as the number of PMT photoelectrons per keV of deposited energy. The sensitivity amounted to 8.7, 15.1 and 16.1 pe/keV at 22-25, 60 and 88 keV respectively. One can see that the sensitivity degrades with the energy decrease. Such an energy dependence in fact rather precisely reflects the reduction of the X-ray ionization yield in liquid Ar with the energy decrease observed elsewhere \cite{XRayYield}. It was described there by the following dependence of the relative ionization yield ($n_e/N_i$) on the energy ($E$): at a drift field of 0.6 kV/cm, $n_e/N_i = 0.54/(1+41/E[keV])$=0.195, 0.320 and 0.368 at 23, 60 and 88 keV respectively.

The fact that the detector yield is proportional (within the experimental uncertainties, of about 5\% \cite{XRayYield}) to the X-ray ionization yield in liquid Ar allows one to estimate, inter alia, the lower limit of the electron lifetime in the liquid, using the fact that the electron drift path in liquid Ar is larger for softer X-rays than that for harder X-rays. The electron lifetime estimated this way exceeds 130 $\mu$s.

\begin{figure}[hbt]
\centering
\includegraphics[width=\columnwidth,keepaspectratio]{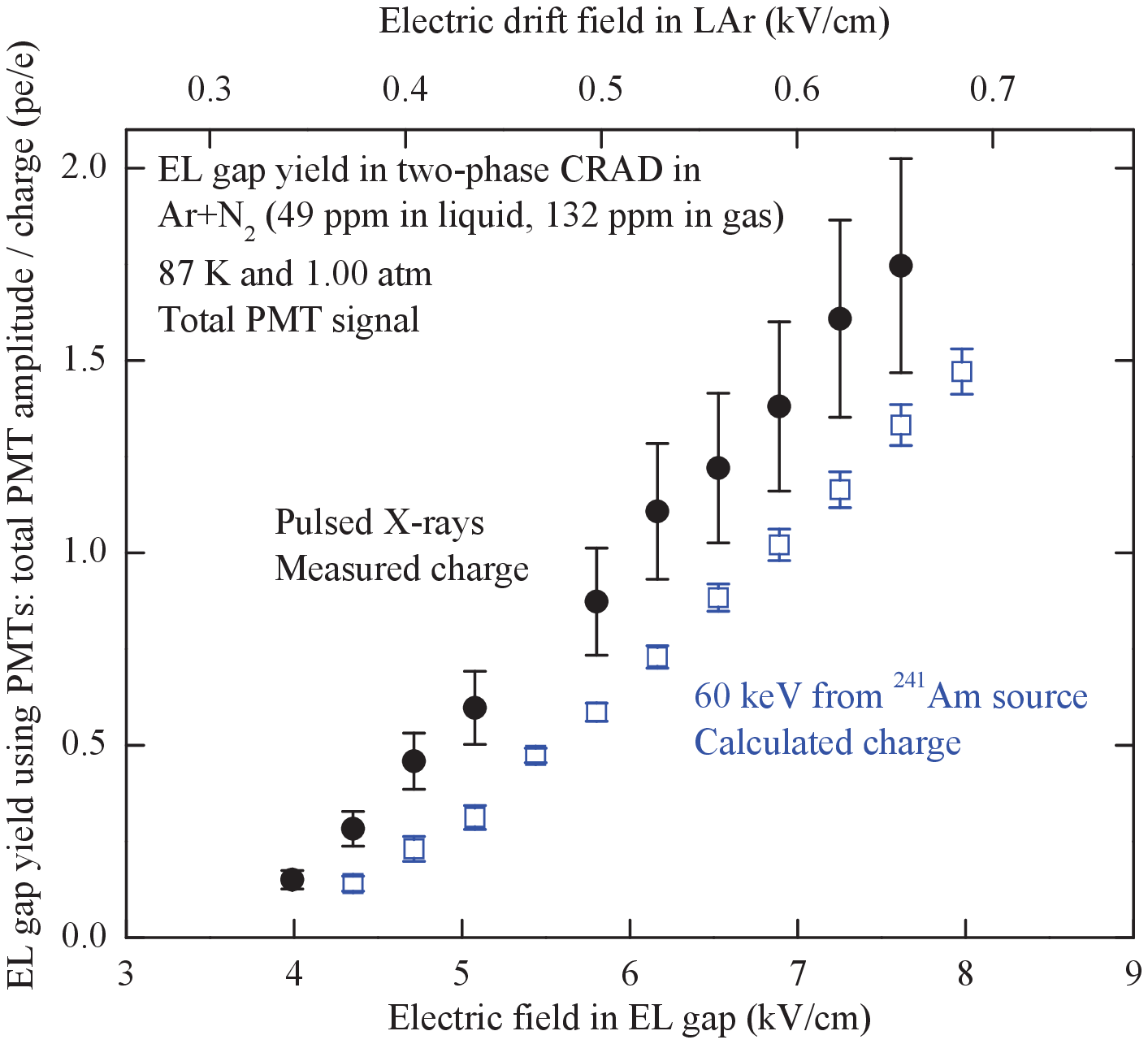}
\caption{EL gap yield measured using PMT signals as a function of the electric field in the EL gap for X-rays from the pulsed X-ray tube, using the measured charge, and for X-rays from $^{241}$Am sources, using the calculated charge. The electric drift field in liquid Ar is also shown on the top axis.}
\label{ELGapYield}
\end{figure}

Another estimate of the lifetime can be obtained from Fig.~\ref{ELGapYield} showing the EL gap yield as a function of the electric field in the gap for X-rays from the pulsed X-ray tube, using the measured charge, and from the $^{241}$Am source, using the calculated charge. In the latter case the ionization charge  arrived to the THGEM1 (anode) was too small to be measured. Consequently it was calculated theoretically by interpolation of the X-ray ionization yield dependence on energy at different electric fields using the data of ref. \cite{XRayYield}. One can see from Fig.~\ref{ELGapYield} that the EL gap yield for 60 keV X-rays is systematically reduced compared to that of pulsed X-rays. This reduction is due to the combined effect of the electron transmittance through the THGEM0 and electron lifetime in the liquid, not taking into account in the charge calculation. Hence we obtain the lower limit on the electron lifetime of 70 $\mu$s. Finally the electron lifetime in the liquid averaged over the two estimates exceeds 100 $\mu$s.

In addition, one can conclude from Fig.~\ref{ELGapYield} that the EL gap yield can be as high as 1.5 pe/e at an electric field in the gap of 7 kV/cm. This is a factor of 3 higher than expected in "pure Ar" approach and is due to the fraction of the N$_2$ emission spectrum recorded directly, i.e. escaping re-emission in the WLS film and thus having a considerably higher (by a factor of 20) light collection efficiency \cite{CRADPropEL}.

\section{Conclusions}

Following our first work in the field \cite{CRADPropEL}, we continued to study the performance of the two-phase Cryogenic Avalanche Detector (CRAD) with electroluminescence (EL) gap, operated in argon doped with a minor (49$\pm$7 ppm) admixture of nitrogen. The results of the measurements of the N$_2$ content, detector sensitivity to X-ray-induced (electron-equivalent recoil) signals, EL gap yield and electron lifetime in the liquid were presented. The detector  sensitivity at a drift field in liquid Ar of about 0.6 kV/cm was measured to be 9 and 16 pe/keV at 23 and 88 keV respectively. The EL gap yield may reach 1.5 pe per drifting electron at an electric field in the gap of 7 kV/cm. The results obtained pave the way to the development of N$_2$-doped two-phase Ar detectors with enhanced sensitivity to the S2 signal. Such detectors are relevant in the field of argon detectors for dark matter search and low energy neutrino detection.

\section{Acknowledgements}

This study was composed of two parts. The first part (section 2) was supported by Russian Foundation for Basic Research (grant 15-02-01821). The second part (section 3) was supported by Russian Science Foundation (project N 14-50-00080). This work was done in the frame of DarkSide20k collaboration.



\end{document}